# Medical Documents Classification Based on the Domain Ontology MeSH

Zakaria Elberrichi      Belaggoun Amel      Taibi Malika

Faculty of Sciences Engineering, University Djillali Liabes of the Sidi belabbes, Algeria

**Abstract:** *This paper addresses the problem of classifying web documents using domain ontology. Our goal is to provide a method for improving the classification of medical documents by exploiting the MeSH thesaurus (Medical Subject Headings) which will allow us to generate a new representation based on concepts. This approach was tested with two well-known data mining algorithms C4.5 and KNN, and a comparison was made with the usual representation using stems. The enrichment of vectors using the concepts and the hyperonyms drawn from the domain ontology has significantly boosted their representation, something essential for good classification. The results of our experiments on the benchmark biomedical collection Ohsumed confirm the importance of the approach by a very significant improvement in the performance of the ontology-based classification compared to the classical representation (Stems) by 30%.*

## 1. Introduction

The intensive use of the Web has led to an explosion of available data, which is wonderful, but unfortunately the side effect is a difficult access to relevant information scattered all over the web. Nowadays, different techniques are developed in the hope to automatically allow a better access to relevant information. These techniques which form a major component of the future Semantic Web require a new formalization of the content and the addition of a semantic description generally performed by metadata. Ontologies, one of the models of knowledge representation most commonly used, address this issue. Simply speaking, they organize knowledge based on the field of application and consist of concepts linked by relations. We will try to evaluate the effect of a conceptual representation of medical document on their automatic classification. The rest of this paper is structured as follows. Related works are presented in Section 2. The structure of our approach is presented in Section 3 with its steps and particularities. The benchmark Ohsumed and the medical ontology Mesh used to test and analyze our approach compared to the bag of stems approach are presented in Section 4. Section 5 contains the experimental results and our comprehension of these results. The conclusion and future works are presented in Section 6.

## 2. Related Works

Document representations for classification are typically based on the classical approach bag-of-words.

However, in recent years, researchers have tried to improve document representation by using conceptual representation. One approach is based on the use of ontologies.

The representation of text as a bag-of-words has been disadvantaged by the ignorance of any relationship between the terms thus the importance of the work of Amine [1] proposing the integration of an ontology (WordNet) to improve the process of clustering text documents.

In recent years, the work of Guyot [4] has helped to show that the use of ontologies in text categorization is a promising way.

Litvak and al. in [6] propose a method of classification of the multilingual documents Web by using a multilingual ontology for the conceptual representation of the documents.

Sanchez and al. in [9] offer the opposite approach to information extraction from web documents, and creates an ontology based on statistical and linguistic methods in a given field. The basic idea is: a good ontology design requires a strong semantics, which implies a relevant and meaningful classification of web documents.

Mu-Hee and al. in [7] embody an approach to automatically classify web documents using domain ontology. Without the use of learning algorithms, or a learning base.

## 3. A conceptual Representation Approach

A main problem to solve for a good classification of documents (texts) is: How to represent texts in order to facilitate their processing, and keep only useful information for the classification? The most widely used representation in this area is the bag of words representation. Much work has been proposed to overcome the limits of this representation. In our approach, we propose a method that uses concepts, this will allow while enriching the representation vector, to reduce its dimension. This, we hope, will give us two crucial advantages for our text classification. This will be done in two ways, for the sake of comparison:

- First, mapping the terms into concepts, having chosen a strategy of matching and disambiguation for an initial enrichment of the representation vector.
- Then, a second enrichment by adding hyperonyms to the representation vector.

### 3.1 The Preprocessing of the Texts

This phase will start with a cleaning process since, the data used for classification are Web documents that contain html tags and images or other noise sources to be deleted, leaving only place to text. To avoid that the various version of the same word will be considered as different words and to keep only the most significant word, we need to perform what we call a preprocessing.

The use of words and stemma for the representation of text requires preprocessing so that the classification is as efficient as possible, and for a better relevance of the information. Indeed, many words provide little (if any) information on the significance of the document.

These are usually called stop words. Another preprocessing task is called Stemming; it simplifies the representation vectors of the texts while increasing their informativeness, by replacing the words by their roots.

1. Stops words: Stops words are the words that have a low weight in the meaning of the texts and are often very common. Their removal during the preprocessing will reduce the size of the texts and will subsequently reduce also the time of the classification process. There are lists of these words for almost every language.

2. Stemming: The method presented by Porter in 1980 for English was the first one. It was used to group words of the same root. It allows the classification process to mimic what a human being do naturally when reading for example a text that contains words with the same roots: for example, if he reads the words walk, walker, walking, he naturally deduce that the document strongly suggests the theme of walking. A processing algorithm on this document without stemming, consider each word by itself.

Whereas, a run of Porter's algorithm will give us one word credited with three hits, so a single issue with greater importance. This algorithm consists of a series of rules that consider how words are actually made in the vocabulary of a language, to determine the common roots. Porter's algorithm is adapted to the language of Shakespeare, but due to its success, it has been adapted to other languages. Their development requires the imperative involvement of linguists. The following figure 1 shows a representation vector cleaned and stemmed.

| ……… | Infect | Clinic | Hemiplegia | ……… |

Figure 1. Example of a representation vector for a text from the Ohsumed benchmark.

### 3.2 The Mapping of the Terms into Concepts

The process of mapping terms into concepts is illustrated with an example shown in Figure 2.

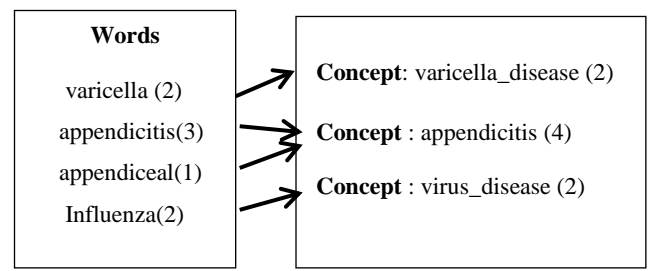

Figure 2. Example of mapping terms into concepts

The words are mapped into their related concepts using the ontology. For example, the two words appendicitis (2) and appendiceal (1) are mapped in the concept appendicitis and the term frequencies of these two words are added in the concept frequency.

From this point, theoretically three strategies for adding or replacing terms by concepts can be distinguished:

- *Add Concept :* This strategy extends each term vector $t_d$ by new entries from MeSH concepts C appearing in the texts set. Thus, the vector $t_d$ will be replaced by the concatenation of $t_d$ and $c_d$. where $c_d = (cf(d,c_1),\ldots,cf(d,c_k))$ the concept vector with $k = |C|$ and $cf(d,c)$ ) denotes the frequency that a concept $c \in C$ appears in a text d. The terms, which also appear in MeSH as a concept, will be accounted for at least twice in the new vector representation; once in the old term vector $t_d$ and at least once in the concept vector $c_d$

- *Replace Terms by Concepts:* This strategy is similar to the first strategy; the only difference lies in the fact that it avoids the duplication of the terms in the new representation; i.e., the terms which appear in Mesh will be taken into account only in the concept vector. The vector of the terms will thus contain only the terms, which do not appear in MeSH.

- *Concept Vector Only:* This strategy differs from the second strategy by the fact that it excludes all the terms from the new representation including the terms, which do not appear in Mesh; $c_d$ is used to represent the category.

### 3.3 The Strategies for Disambiguation

The assignment of terms to concepts is ambiguous, since we deal with natural language. One word may have several meanings and thus one word may be mapped into several concepts. In this case, we need to determine which meaning is being used, which is the problem of sense disambiguation. WSD is considered an AI-complete problem, that is, a task whose solution is at least as hard as the most difficult problems in artificial intelligence [8]. Since a sophisticated solution for sense disambiguation is often impractical and complex [2], we will consider only two simple disambiguation strategies.

- *All Concepts:* This strategy considers all proposed concepts as the appropriate ones for augmenting the text representation. This strategy is based on the assumption that texts contain central themes that in our cases will be indicated by certain concepts having higher weights. These concepts will automatically emerge, but the dimensionality will increase. In this case, the concept frequencies are calculated as follows:

$$cf(d,c) = tf\{d, t \in T \mid c \in (ref_c(t))\}$$

- *First Concept:* This strategy considers only the most often used sense of the word as the most appropriate concept. This strategy is based on the assumption that the used ontology returns an ordered list of concepts in which more common meanings are listed before less common ones. This is the case for most Ontologies. In this case, the concept frequencies are calculated as follows:

$$cf(d,c) = tf\{d, t \in T \mid first(ref_c(t)) = c\}$$

### 3.4 Using Hyperonyms

If concepts are used to represent texts, the relations between concepts can play a key role in capturing the ideas in these texts. Recent researches show that simply replacing the terms by concepts, without considering the relations, does not have a significant improvement and sometime even perform worse than terms [2]. For this purpose, we have considered the hyperonyms relation between concepts by adding to the concept frequency of each concept in a text the frequencies that their hyponyms appears. Then the frequencies of the concept vector part are updated in the following way:

$$cf'(d,c) = \sum_{b \in H(c)} cf(d,b)$$

Where H(c) gives for a given concept c its hyponyms.

### 3.5 Descriptors Selection and Reduction

The mapping operation is performed on the learning corpus for each document and each document will be represented by a vector whose descriptors are the concepts of the ontology. Every concept is associated with the frequency of appearance in the learning corpus of the category.
This selection is to choose for each category the descriptors that characterize it best compared to the other categories.

A weighting is used to represent the importance of the term in a category. The number of occurrences in the category is the easiest way to calculate this value, but it is not very satisfactory in the sense that it does not take into account its importance for the other categories.

A better and more widely used weighting is known as the TF-IDF. It was introduced for the vector model, it means:

«*term frequency* »* « *Inverse document frequency* »

"TF: Term Frequency " is simply the number of occurrences of the term in the relevant category, the "IDF: Inverse Document Frequency "is the inverse of the total number of category divided by the number of categories containing the term we want to weigh, nbr-category is the number of categories is the formula:

$$TFIDF(c_i, w_j) = TF(c_i, w_j) * \log\left(\frac{nbr\_category}{DF(w_j)}\right)$$

Selection techniques for dimensionality reduction take as input a set of features and output a subset of these features, which are more relevant for discriminating among categories. Controlling the dimensionality of the vector space is essential for two reasons. The complexity of many learning algorithms depends crucially not only on the number of learning examples but also on the number of features. Thus, reducing the number of index terms may be necessary to make these algorithms tractable. Also, although more features can be assumed to carry more information and should, thus, lead to more accurate classifiers, a larger number of features with possibly many of them being irrelevant may actually hinder a learning algorithm constructing a classifier.

For our approach, a feature selection technique is necessary in order to reduce the big dimensionality. For this purpose we used the Chi-Square Statistic for feature selection. The χ2 statistic measures the degree of association between a term and a category. Its application is based on the assumption that a term whose frequency strongly depends on the category in which it occurs will be more useful for discriminating it among other categories. For the purpose of dimensionality reduction, terms with small χ2 values are discarded. The χ2 multivariate is a supervised method allowing the selection of terms by taking into account not only their frequencies in each category but also the interaction of the terms between them and the interactions between the terms and the categories. The principle consists in extracting K better features characterizing best the category compared to the others, this for each category.

An arithmetically simpler way of computing χ2 is the following:

$$X^2(D, t, c) = \frac{(N_{11} + N_{10} + N_{01} + N_{00}) * (N_{11}N_{00} - N_{10} + N_{01})^2}{(N_{11} + N_{01}) * (N_{11} + N_{10}) * (N_{10} + N_{00}) * (N_{01} + N_{00})}$$

Where**:**

$N_{tc}$ : Number of document such as t, c {0.1}

$N_{11}$: The number of documents containing the term and in the category

$N_{10}$: The number of documents containing the term and not in the category.

$N_{01}$: The number of document which does not contain the term and are in the category.

$N_{00}$: The number of document which does not contain the term and not in the category.

The principal characteristics of this method are:

• It is supervised because it is based on the information brought by the category

• It is a multivariate method because it evaluates the role of the feature with considering the other features.

• It considers interactions between features and categories.

• In spite of its sophistication, it remains of linear complexity in terms number**.**

### 3.6 Classification of the Texts

Once, the preprocessing is done and the concept's representation is performed, as in any supervised classification we build the model using the matrix formed by the concept vectors, and a machine learning algorithm. We will evaluate our approach on two of the most popular classification algorithms the C4.5 and the KNN. Since they are very popular, and since our goal is more te test of the conceptual presentation, they don't need any presentation.

Once, the model is built, to classify a new text, we expose its vector of concepts, generated in the same way explained above, to the model created in the learning phase, to find the appropriate class. The figure 3 presents the details of the whole method.

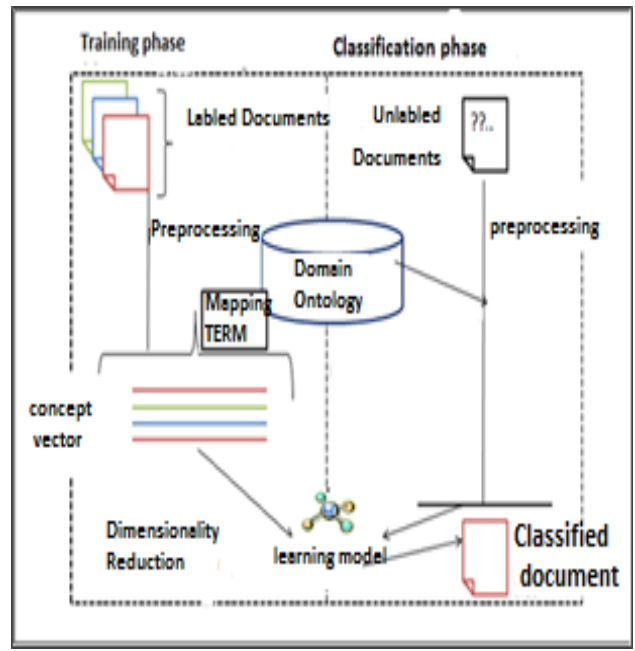

Figure 3. A conceptual representation approach for document classification.

## 4. The evaluation of the Approach

### 4.1 The Ohsumed Collection

We use the collection OHSUMED proposed in the framework of the Task-Filtering TREC9 in 2000, which is made of titles and / or summaries de 270 medical journals published between 1987-1991 [5]. A document contains six fields: title (.T), summary (.W), MeSH indexed concepts (.M), author (.A), source (.S), and publication (.P).

Table 1. Details of Ohsumed categories.

| Category | #Docs |
|---|---|
| **Bacterial Infections and Mycoses** | 2540 |
| **Virus Diseases** | 1171 |
| Parasitic Diseases | 427 |
| Neoplasms | 6327 |
| Musculoskeletal Diseases | 1678 |
| **Digestive System Diseases** | 2989 |
| **Stomatognathic Diseases** | 526 |
| **Respiratory Tract Diseases** | 2589 |
| Otorhinolaryngologic Diseases | 715 |
| **Nervous System Diseases** | 3851 |
| **Eye Diseases** | 998 |
| Urologic and Male Genital Diseases | 2518 |
| Female Genital Diseases and Pregnancy Complications | 1623 |
| Cardiovascular Diseases | 6102 |
| Hemic and Lymphatic Diseases | 1277 |
| Neonatal Diseases and Abnormalities | 1086 |
| Skin and Connective Tissue Diseases | 1617 |
| Nutritional and Metabolic Diseases | 1919 |
| Endocrine Diseases | 865 |
| Immunologic Diseases | 3116 |
| Disorders of Environmental Origin | 2933 |
| Animal Diseases | 506 |
| Pathological Conditions, Signs and Symptoms | 9611 |

### 4.2 The Evaluation Method

Experimental results reported in this section are based on the so-called "F-measure", which is the harmonic mean of precision and recall.

$$F = \frac{2 * recall * precision}{recall + precision}$$

In the above formula, precision and recall are two standard measures widely used in text categorization literature to evaluate the algorithm's effectiveness on a given category where

$$precision = \frac{True\ positive}{True\ positive + false\ positive}$$

$$recall = \frac{True\ positive}{True\ positive + false\ negative}$$

### 4.3 The Ontology MeSH

We used the biomedical thesaurus reference developed by NLM in the U.S.A. The MeSH thesaurus (Medical Subject Headings) is a tool created by the National Library of Medicine (NLM). It is used for indexing and for medical information retrieval. The first version appeared in 1954 as the Subject Heading Authority List. It was published as the Medical Subject Headings in 1963 and contained in this edition, 5700 descriptors.

Faced with the growing number of medical resources to be managed by medical librarians, the NLM launched the project called at this time MEDLARS to automate the indexing and retrieval of medical resources. Since, MeSH has evolved. It had 25,588 descriptors in its 2010. The MeSH descriptors are organized into 16 categories: category A for anatomic terms, category B for organisms, the category C for diseases, etc.. Each category is subdivided into subcategories. Within each category, the descriptors are hierarchically structured from general to specific, with a level of maximum depth of 11.

MeSH is the thesaurus of controlled vocabulary for indexing resources MEDLINE20 bibliographic database. It is also used by portals as indexing and cataloging of medical resources, Health On the Net and CisMeF. INSERM maintains a French version of MeSH

## 5. Results

In order to demonstrate the utility of using MeSH in the classification, we tested the proposed approach on our database (Ohsumed) with two data mining algorithms C4.5 and KNN, respectively, the K nearest neighbor and the decision trees, which gave their evidence of success in the classification of textual documents. The following table summarizes the results of our approach compared with the stems representation. The approach is tested on eight categories of the Ohsumed corpus (bold in table 1). Table 2 presents the results (F-measures) obtained with the mode 5-fold cross-validation and the Chi2 reduction technique

Table 2. F-measure for concepts and Stem (8 categories)

| Descriptors | Concepts | | Concepts + Hyperonym | | Stems | |
|---|---|---|---|---|---|---|
| Algorithms | KNN | C4.5 | KNN | C4.5 | KNN | C4.5 |
| C1 | 0.962 | 0.959 | 0.961 | 0.936 | 0.450 | 0.511 |
| C2 | 0.953 | 0.919 | 0.957 | 0.928 | 0.667 | 0.623 |
| C3 | 0.927 | 0.705 | 0.938 | 0.936 | 0.581 | 0.629 |
| C4 | 0.926 | 0.936 | 0.95 | 0.887 | 0.629 | 0.5 |
| C5 | 0.933 | 0.954 | 0.82 | 0.951 | 0.69 | 0.421 |
| C6 | 0.942 | 0.935 | 0.958 | 0.939 | 0.545 | 0.427 |
| C7 | 0.954 | 0.943 | 0.959 | 0.949 | 0.5 | 0.468 |
| C8 | 0.598 | 0.672 | 0.627 | 0.497 | 0.606 | 0.487 |
| **AvG** | **0.919** | **0.89** | **0.923** | **0.908** | **0.601** | **0.531** |

Given the results, we can say that the representation based on ontology provides clearly better performance. A significant performance upgrading of 30% is an unexpected response and very outstanding news for any researcher who implements a hypothetical approach.

The enrichment of the representation vector by hyperonyms in addition to related concepts is a good idea, since the performance gain is even better. The choice of the algorithms was justified, the KNN were proven effective with outstanding results. One reason for this, in our opinion is their compatibility with the CHI 2 reduction technique.

## 6. Conclusion and Future Works

The main objective of our approach was to improve the classification process using domain ontology. Our method was tested on a specific area, the medical field, given its importance and the interest it raises currently in the data mining community.

The approach was tested on a benchmark corpus with two popular algorithms the KNN and the C4.5. We did it three times. First, with the classical Stems representation, for the sake of comparison. Then the two proposed ones. Primary, using the related concepts for document's representation, and secondly, using the concepts and the hyperonyms, of course provided by MeSH. The results (+30%) show the success of our approach.

Therefore, these results prove that document classification in a particular area supported by ontology of this domain is without a doubt a promising method.

For future works, we see many things that remain to be exploited. Firstly, we can go further than one level (Hyperonymy) in the domain ontology MeSH, trying to generalize to the maximum possible, and see the impact on the performance. Secondly, we can study the classification of multilingual medical documents using the same conceptual approach based on a multilingual MeSH ontology [3].

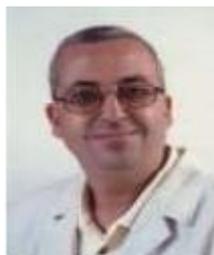
Zakaria Elberrichi is senior lecturer in machine learning and data mining at the Djillali Liabes University of Sidi belabbes, Algeria. He holds PhD degree in computer science from the same university. In addition he is a member and the team leader of EEDIS (Evolutionary *Engineering & Distributed Information Systems*) Laboratory. Dr. Elberrichi has more than 22 years of experience including extensive project management experience in planning and leading a range of data mining-related projects. Dr. Elberrichi supervises eight PhD students in Web Semantic Mining, GIS Systems, E-learning, Social Network Mining, Biometrics, Arabic Text Mining, and Biomedical Mining. He also leads and teaches modules at MSc levels in computer science.

Belaggoun Amel and Taibi Malika received their Master with thesis degree from the computer science department at the UDL University with honor in 2011, and are currently doctorate students and members of the research team intelligent web mining.